# THE ANALYSIS OF THE IFPUG METHOD SENSITIVITY

Ramón Asensio Monge, Francisco Sanchìs Marco and Fernando Torre Cervigòn

**Abstract-** J. Albrecht`s Function Point Analysis (FPA) is a method to determine the functional size of software products. An organization called International Function Point Users Group (IPFUG), considers the FPA as a standard in the software functional size measurement. The Albrecht´s method is followed by IPFUG method which includes some modifications in order to improve it. A limitation of the method refers to the fact that FPA is not sensitive enough to differentiate the functional size in small enhancements. That affects the productivity analysis, where the software product functional size is required. To provide more power to the functional size measurement, A. Abran, M. Maya and H. Nguyeckim have proposed some modifications to improve it. The IPFUG v 4.1 method which includes these modifications is named IFPUG v 4.1 extended. In this work we set the conditions to delimiting granular from non granular functions and we calculate the static calibration and sensitivity graphs for the measurements of a set of projects with a high percentage of granular functions, all of then measured with the IFPUG v 4.1 method and the IFPUG v 4.1 extended. Finally, we introduce a statistic analysis in order to determine whether significant differences exist between both methods or not.

## 1 INTRODUCTION

A limitation of FPA arises from the classification of functional complexity for each function type. FPA considers three levels of complexity: low, medium or high, which, according to C.R. Simmons [1] represent an excessive simplification. Function points are mostly used to measure medium and big developments or project enhancements. Nevertheless, the vast majority of maintenance requests correspond with a low complexity and the measurements have the same results.

According to M. Maya and A. Abram [2] [3] the present structure of FPA is inadequate in these cases.

The FPA has not enough sensitivity to measure the functional size of small enhancement projects, considering the enhancements as the changes that add or modify the functionality, due to the changes of data or process requirements. The same problems appear when we measure development projects with high percentage of granular functions.

In this work, we consider granular functions as those that fulfil certain conditions (deeply studied in point 5) connected to the number of record element types RETs, file types referenced FTRs or data elements types DETs, that suppose a greater or smaller functionality in the corresponding function type. Sensitivity is considered as the functionality variation related to the variation of the previously mentioned variables. This is an essential point in the research of measurement methods.

## 2. IFPUG V. 4.1 METHOD (FP4.1)

In January of 1999, the 4.1 version of the IFPUG counting practices manual was published [4]. New rules are incorporated, others are changed to solve not previously studed situations, and new examples and indications are added, so that they help to understand the method.

The method considers five function types:

- Internal Logic File (ILF): It is a user identifiable group of logically related data or control information maintained within the boundary of the application. The primary intent of an ILF is to hold data maintained through one or more elementary processes of the application being counted.

- External Interface File (EIF): It is a user identifiable group of logically related data or control information referenced by the application, but maintained within the boundary of another application. The primary intent of an EIF is to hold data referenced through one or more elementary processes within the boundary of the application counted.

- External Input (EI): It is an elementary process that processes data or control information that comes from outside the application boundary. The primary intent of an EI is to maintain one or more ILFs and/or to alter the behaviour of the system.

- External Output (EO): It is an elementary process that sends data or control information outside the application boundary. The primary intent of an external output is to present information to a user through processing logic other than, or in addition to, the retrieval of data or control information. The processing logic must contain al least one mathematical formula or calculation, or create derived data. An EO may also maintain one or more ILFs and/or alter the behaviour of the system.

- External Inquiry (EQ): It is a elementary process that sends data or control information outside the application boundary. The primary intent of an EQ is to present information to a user, through the retrieval of data or control information from an ILF or EIF. The processing logic contains no mathematical formulas or calculations, and it creates no derived data. It does not maintain any ILF and it can not modify the system behaviour.

The complexity of ILFs and EIFs is determined by the number of RETs and DETs. A data element type (DET) is each unique user recognizable, non-repeated field maintained in or retrieved from the ILF or EIF through the execution of an elementary process. A record element type (RET) is a subgroup of data elements, recognizable by the user within an ILF or EIF.

Rate the functional complexity of the ILFs and EIFs using the table 1.

Table 1: ILFs and EIFs functional complexity

|  | 1 a 19 DETs | 20 a 50 DETs | 51 o más DETs |
|---|---|---|---|
| 1 RET | Low | Low | Average |
| 2 a 5 RETs | Low | Average | High |
| 6 o más RETs | Average | High | High |

The complexity of EIs, EOs and EQs is determined by the number of FTRs and DETs. A file type referenced (FTR) is an ILF read or maintained by a process function type, or an EIF read by a process function type. A data element type (DET) is a unique user recognizable, non-repeated field.

The tables 2 and 3 can be used to rate the functional complexity for EIs, EOs and EQs.

Table 2: EIs Functional Complexity

|  | 1 a 4 DETs | 5 a 15 DETs | 16 o más DETs |
|---|---|---|---|
| 0 ó 1 FTR | Low | Low | Average |
| 2 FTRs | Low | Average | High |
| 3 o más FTRs | Average | High | High |

Table 3: EOs and EQs Functional Complexity

|  | 1 a 5 DETs | 6 a 19 DETs | 20 o más DETs |
|---|---|---|---|
| 0 ó 1 FTR | Low | Low | Average |
| 2 o 3 FTRs | Low | Average | High |
| 4 o más FTRs | Average | High | High |

The table 4 can be used to translate the ILF, EIF, EI, EO or EQ complexity to unadjusted function points.

Table 4: Function point count per function type

| Function Type | Functional Complexity Rating | | |
|---|---|---|---|
|  | Low | Average | High |
| ILF | 7 | 10 | 15 |
| EIF | 5 | 7 | 10 |
| EI | 3 | 4 | 6 |
| EO | 4 | 5 | 7 |
| EQ | 3 | 4 | 6 |

## 3. IFPUG v. 4.1 EXTENDED METHOD (EFP4.1)

When the record element types (RETs) and data element types (DETs) in each ILF or EIF are very low, the corresponding ILF or EIF complexity is low. These intervals are shaded in table 1. Similarly, when the file types referenced (FTRs) and the data element types (DETs) in EIs, EOs and EQs are very low, the internal functional complexity is low, as it can be observed shaded in tables 2 and 3. This happens in a high percentage of enhancement projects [3] and in development projects with a high percentage of granular functions.

In these cases, in order to differentiate the functionality given to each function type, A. Abran, M. Maya and H. Nguyenkim [5] [6] proposed as a solution a subdivision of the first intervals in intermediate subintervals. The count of function points for each function type is as follows:

- ILFs: From 2 intervals with weight 7, to 25 intervals with weights from 1 to 7, as it appears in table 5

- EIFs: From 2 intervals with weight 5, to 25 intervals with weights from 1 to 5, as it appears in table 6

Table 5, ILFs

|         | 1-3 DETs | 4-6 DETs | 7-9 DETs | 10-14 DETs | 15-19 DETs | 20-50 DETs | 51+ DETs |
|---------|----------|----------|----------|------------|------------|------------|----------|
| 1 RET   | 1        | 1        | 2        | 3          | 5          | 7          | 10       |
| 2 RETs  | 1        | 2        | 3        | 5          | 7          | 10         | 15       |
| 3 RETs  | 2        | 3        | 5        | 7          | 7          |            |          |
| 4 RETs  | 3        | 5        | 7        | 7          | 7          |            |          |
| 5 RETs  | 5        | 7        | 7        | 7          | 7          |            |          |
| 6+ RETs | 10       |          |          |            |            | 15         | 15       |

Table 6. EIFs

|         | 1-3 DETs | 4-6 DETs | 7-9 DETs | 10-14 DETs | 15-19 DETs | 20-50 DETs | 51+ DETs |
|---------|----------|----------|----------|------------|------------|------------|----------|
| 1 RET   | 1        | 1        | 2        | 3          | 4          | 5          | 7        |
| 2 RETs  | 1        | 2        | 3        | 4          | 5          | 7          | 10       |
| 3 RETs  | 2        | 3        | 4        | 5          | 5          |            |          |
| 4 RETs  | 3        | 4        | 5        | 5          | 5          |            |          |
| 5 RETs  | 4        | 5        | 5        | 5          | 5          |            |          |
| 6+ RETs | 7        |          |          |            |            | 10         | 10       |

- EIs: From 3 intervals with weight 3, to 8 intervals with weights from 0,5 to 3, as it appears in table 7
- EOs: From 3 intervals with weight 4, to 14 intervals with weights from 0,5 to 4, as it appears in table 8
- EQs: From 3 intervals with weight 3, to 8 intervals with weights from 0,5 to 3, as it appears in table 9.

Table 7, EIs

|         | 1-2 DETs | 3-4 DETs | 5-6 DETs | 7-8 DETs | 9-11 DETs | 12-15 DETs | 16+ DETs |
|---------|----------|----------|----------|----------|-----------|------------|----------|
| 0 ó 1 FTR | 0,5    | 1        | 1,5      | 2        | 2,5       | 3          | 4        |
| 2 FTRs  | 2        | 3        | 4        |          |           |            | 6        |
| 3+ FTRs | 4        |          | 6        |          |           |            | 6        |

Table 8, EOs

|         | 1 DET | 2-3 DETs | 4-5 DETs | 6-7 DETs | 8-9 DETs | 10-12 DETs | 13-15 DETs | 16-19 DETs | 20+ DETs |
|---------|-------|----------|----------|----------|----------|------------|------------|------------|----------|
| 0 ó 1 FTR | 0,5 | 1        | 1,5      | 2        | 2,5      | 3          | 3,5        | 4          | 5        |
| 2 FTRs  | 1     | 1,5      | 2        |          |          |            |            |            |          |
| 3 FTRs  | 1,5   | 2        | 4        |          | 5        |            |            |            | 7        |
| 4+ FTRs |       | 5        |          |          | 7        |            |            |            | 7        |

Table 9, EQs

|         | 1-2 DETs | 3-4 DETs | 5-6 DETs | 7-8 DETs | 9-11 DETs | 12-15 DETs | 16+ DETs |
|---------|----------|----------|----------|----------|-----------|------------|----------|
| 0 ó 1 FTR | 0,5    | 1        | 1,5      | 2        | 2,5       | 3          | 4        |
| 2 FTRs  | 2        | 3        | 4        |          |           |            | 6        |
| 3+ FTRs | 4        |          | 6        |          |           |            | 6        |

This subdivision must provide more power to the measurement method, and, at the same time, it will allow a better analysis in productivity studies.

We call IFPUG v. 4.1 Extended Method to a revision[1] of the IFPUG v. 4.1 that incorporates A. Abran, M. Maya and Nguyenkim`s previous tables in order to determine the unadjusted function point count.

---

[1] Method modification to improve it, but without varying the application domain.

# 4. SENSITIVITY OF EACH METHOD

Calculating the sensitivity of one method in relation to its correspondent extended method will imply the determination of static calibration and sensitivity graphs for each of them. The sensitivity of a method in each point is given by the slope of static calibration graph in that point.

The concept of static calibration refers to the fact that every method´s inputs are fixed to constant values, except one, which can vary into an interval. This procedure can be repeated changing the fixed variables. If global effects were to be studied, instead of individual effects, the calibration procedure would specify the variety of several inputs simultaneously.

In this work the static calibration graphs represent the functionality increase in relation to the increasing number of RETs, FTRs and DETs. The sensitivity graphs shows the relation between the increase of functionality and the increase of number of RETs, FTRs and DETs.

To investigate the sensitivity of a method requires several measurements of the same projects with the same method, varying one or several input variables into a range of values and holding fixed the others.

# 5. SENSITIVITY EMPIRICAL EVALUATIONS

To empirically evaluate sensitivity of the IFPUGs v. 4.1 and IFPUG v 4.1 extended methods, 30 management information system projects have been selected. These projects correspond to Administration, Finances, Services and Industry sectors.

The documentation level needed was the Requirements Analysis, corresponding to level one in Rudolph´s classification [7].

Four raters were selected and organised in the following subgroups: M1-M2, M1-M3, M1-M4, M2-M3, M2-M4, M3-M4. Thus, each rater measured 15 projects and each project was measured by 2 raters.

The projects distribution are shown in table 10.

Table 10, Projects assigned to raters

| RATERS | ORGANIZATION / PROJECTS ||||||||||||||||
|---|---|---|---|---|---|---|---|---|---|---|---|---|---|---|---|---|
| | ORGANIZAT. 1 ||| ORGANIZAT. 2 ||| ORGANIZAT. 3 ||| ORGANIZAT. 4 |||| ORGANIZAT. 5 |||
| M1 | P1 | P2 | P3 | P7 | P8 | P9 | P13 | P14 | P15 | P19 | P20 | P21 | P25 | P26 | P27 | |
| M2 | p1 | p4 | p5 | p7 | p10 | p11 | p13 | p16 | p17 | p19 | p22 | p23 | p25 | p28 | p29 | |
| M3 | p2 | p4 | p6 | p8 | p10 | | p12 | p14 | p16 | p18 | p20 | p22 | p24 | p26 | p28 | p30 |
| M4 | p3 | p5 | p6 | p9 | p11 | | p12 | p15 | p17 | p18 | p21 | p23 | p24 | p27 | p29 | p30 |

A set of computer programs was necessary to facilitate the measurement, simulation and statistic analysis. The programming language used was Fortran and SPSS statistic package, since they offer an adequate solution to the statistic analysis needed in this work.

**Measurement results**

The count number of function points for each project measured by FP4.1 and EFP4.1 methods are shown in table 11.

Table 11. Count of function points per project

| Projects | FP4.1 Method | EFP4.1 Method |
|---|---|---|
| 1 | 291,00 | 210,00 |
| 2 | 275,00 | 224,75 |
| 3 | 173,00 | 155,00 |
| 4 | 218,00 | 171,00 |
| 5 | 139,50 | 89,00 |
| 6 | 229,50 | 148,25 |
| 7 | 252,00 | 238,50 |
| 8 | 374,00 | 342,25 |
| 9 | 221,50 | 205,50 |
| 10 | 272,00 | 218,00 |
| 11 | 188,00 | 170,50 |
| 12 | 436,00 | 320,00 |
| 13 | 277,00 | 198,50 |
| 14 | 451,50 | 344,75 |
| 15 | 548,00 | 433,00 |
| 16 | 256,00 | 201,00 |
| 17 | 352,50 | 188,50 |
| 18 | 441,50 | 323,50 |
| 19 | 345,00 | 248,75 |
| 20 | 268,50 | 195,00 |
| 21 | 110,50 | 88,50 |
| 22 | 135,00 | 107,00 |
| 23 | 256,00 | 213,00 |
| 24 | 187,00 | 160,75 |
| 25 | 263,50 | 161,00 |
| 26 | 408,50 | 348,50 |
| 27 | 664,00 | 448,75 |
| 28 | 256,00 | 148,75 |
| 29 | 116,50 | 93,00 |
| 30 | 242,00 | 138,25 |

The minimum and maximum value, the mean and the standard deviation of the unadjusted function points for each method are shown in table 12.

Table 12. Mean and standard deviation per method

| Method | N | Minimum | Maximum | Mean | Std. Deviation |
|---|---|---|---|---|---|
| FP4.1 | 30 | 110,50 | 664,00 | 288,2833 | 127,4508 |
| EFP4.1 | 30 | 88,50 | 448,75 | 217,7750 | 95,7009 |

To contrast if there are significant differences between the measurement values given by both methods [8] [9] [10], the following steps are needed:

1. Determine the possible influence of the raters in the measurements. The results of the test are shown in table 13. As it can be seen the p-value is >0,05 for both methods and, consequently, the null hypothesis is accepted; therefore, the influence of raters is the same.
2. Eliminate the influence of raters. In order to eliminate the possible influence of raters, we consider for the pair of methods A and B the new variable dab, which is given by the arithmetical mean of the differences between the measurements by each project and method:

$$d1ab = M_{1A} - M_{1B}$$
$$d2ab = M_{2A} - M_{2B}$$
$$dab = ( d1ab + d2ab ) / 2$$

A and B are the two methods FP4.1 and EFP4.1 respectively.

3. Compare the values of the measurements provided by both methods. We would start by testing whether the dab variable has a normal distribution or not. The Kolmogorov-Smirnov test results are shown in table 14.

Table 13, Analysis of variance

| Method | Source | Sum of squares | D.F. | Mean Square | F | p-value |
|---|---|---|---|---|---|---|
| FP4.1 | Between Groups | 8765,750 | 5 | 1753,150 | 1,407 | ,257 |
| | Within Groups | 29904,400 | 24 | 1246,017 | | |
| | Total | 38670,150 | 29 | | | |
| EFP4.1 | Between Groups | 10034,721 | 5 | 2006,944 | 1,547 | ,213 |
| | Within Groups | 31136,200 | 24 | 1297,342 | | |
| | Total | 41170,921 | 29 | | | |

Table 14. Komogorov-Smirnov goodness of fit test

Test distribution - normal

| Variable | N | Normal parameters | | Diferencias más extremas | | | Kolmogorov-Smirnov Z | p-value |
|---|---|---|---|---|---|---|---|---|
| | | Mean | Std. desviation | Absolute | Positive | Negative | | |
| dab | 30 | 70,5083 | 47,7459 | ,127 | ,127 | -,116 | ,697 | ,715 |

Being the p-value greater than 0.05 the dab variable is normally distributed. Consequently, a Student´s t test will be applied. The result is shown in table 15.

Table 15. T-test. Sample value = 0

| Variable | t-value | D.F. | p-value | Paired Differences | 95% CI | |
|---|---|---|---|---|---|---|
| | | | | | Lower | Upper |
| dab | 8,088 | 29 | ,000 | 70,5083 | 52,6797 | 88,3370 |

Being the p-value lower than 0,05 the test shows that there are significant differences between the values of the measurements given by methods: FP4.1 and EFP4.1.

**Granularity**

We call granular functions those that fulfil the conditions shown in table 16, according to the number of RETs and DETs or FTRs and DETs, determined for each function type in accordance with the IFPUG v 4.1 method definitions, rules and indications. Non-granular functions are those not fulfilling.

Table 16. Granularity conditions

| Function Type | Parameters | Granularity conditions |
|---|---|---|
| ILF | RETs DETs | RETs less or equal than 5 and DETs less or equal than 19 |
| EIF | RETs DETs | RETs less or equal than 5 and DETs less or equal than 19 |
| EI | FTRs DETs | FTRs less or equal than 1 and DETs less or equal than 15 or FTRs equal 2 and DETs less or equal than 4 |
| EO | FTRs DETs | FTRs less or equal than 1 and DETs less or equal than 19 or FTRs less or equal than 3 and DETs less or equal than 5 |
| EQ | FTRs DETs | FTRs less or equal than 1 and DETs less or equal than 19 or FTRs less or equal than 3 and DETs less or equal than 5 |

Table 17. Granular functions percent per project [2]

| Projects | ILF | | EIF | | EI | | EO | | EQ | | Granular functions percent |
|---|---|---|---|---|---|---|---|---|---|---|---|
| | Number of granular functions | Number of non granular functions | Number of granular functions | Number of non granular functions | Number of granular functions | Number of non granular functions | Number of granular functions | Number of non granular functions | Number of granular functions | Number of non granular functions | |
| 1 | 6,00 | 3,00 | 13,50 | ,00 | 4,00 | 18,00 | 1,00 | 2,50 | 1,50 | 4,50 | 48% |
| 2 | 6,50 | 1,50 | 4,00 | ,00 | 11,00 | 12,50 | 1,00 | 2,00 | ,00 | 14,50 | 42% |
| 3 | 1,50 | ,50 | 6,00 | ,50 | 1,00 | 13,00 | ,00 | 1,00 | 2,00 | 7,50 | 32% |
| 4 | 4,50 | ,50 | 8,00 | ,00 | 7,50 | 8,00 | ,50 | 1,50 | 1,00 | 11,50 | 50% |
| 5 | ,00 | ,00 | 18,00 | ,50 | ,00 | ,00 | ,00 | 2,50 | ,50 | 4,50 | 71% |
| 6 | 9,00 | 1,00 | 4,00 | ,00 | 11,50 | 11,50 | 2,00 | ,50 | 6,50 | 6,00 | 63% |
| 7 | 4,00 | 5,50 | 1,00 | ,00 | 4,00 | 19,00 | ,00 | 2,50 | 1,50 | 6,00 | 24% |
| 8 | 8,00 | 7,00 | 3,50 | ,50 | 11,00 | 25,00 | ,00 | 3,00 | ,50 | 8,00 | 35% |
| 9 | ,50 | 1,50 | 10,00 | 4,00 | ,00 | 5,00 | ,00 | 13,50 | ,50 | 1,00 | 31% |
| 10 | 10,00 | 2,50 | 4,50 | ,50 | 4,50 | 15,50 | 1,50 | 7,00 | ,00 | ,00 | 45% |
| 11 | 2,00 | 2,00 | 3,00 | ,00 | 2,00 | 20,00 | 1,00 | ,00 | 1,00 | 1,00 | 28% |
| 12 | 13,50 | 4,50 | 4,00 | ,00 | 13,00 | 23,50 | 3,00 | 7,00 | 6,00 | 11,00 | 46% |
| 13 | 7,50 | 2,00 | 3,00 | ,00 | 11,00 | 9,50 | 2,00 | 2,50 | 4,00 | 13,50 | 50% |
| 14 | 12,00 | 3,00 | 9,00 | ,00 | 14,50 | 13,50 | 1,00 | 12,50 | 2,50 | 15,00 | 47% |
| 15 | 12,00 | 5,00 | 11,00 | ,00 | 12,00 | 23,50 | 1,00 | 11,50 | 2,00 | 21,50 | 38% |
| 16 | 2,00 | 3,00 | 8,50 | ,00 | 9,00 | 11,00 | 2,50 | 6,50 | 2,00 | 7,00 | 47% |
| 17 | 18,00 | ,00 | 2,00 | ,00 | 19,50 | 5,00 | 3,50 | 8,00 | 7,00 | 6,50 | 72% |
| 18 | 9,00 | 2,50 | 12,00 | ,00 | 13,00 | 18,00 | 4,50 | 12,50 | 10,50 | 9,50 | 54% |
| 19 | 12,50 | ,50 | 4,00 | ,50 | 16,00 | 16,00 | 3,00 | 11,50 | 2,00 | 4,00 | 54% |
| 20 | 8,00 | 2,00 | 5,00 | ,00 | 6,50 | 16,50 | 4,00 | 5,50 | 1,00 | 3,50 | 47% |
| 21 | ,00 | 1,00 | 6,50 | ,00 | ,00 | 4,50 | ,50 | 5,50 | ,00 | 1,50 | 36% |
| 22 | 1,00 | 3,00 | 5,50 | ,50 | 3,50 | 7,50 | ,50 | ,50 | ,50 | 4,00 | 42% |
| 23 | 4,50 | 2,50 | 9,50 | 2,00 | 2,50 | 12,50 | ,00 | 9,50 | ,00 | ,50 | 38% |
| 24 | 4,50 | 2,50 | 5,00 | ,50 | 2,50 | 9,00 | ,00 | 3,50 | ,00 | 5,50 | 36% |
| 25 | 13,00 | ,00 | ,00 | ,00 | 20,50 | 8,50 | 2,00 | 7,00 | 2,00 | 3,00 | 67% |
| 26 | 5,50 | 3,50 | 8,00 | ,00 | 3,50 | 36,50 | ,00 | 10,50 | ,50 | 2,50 | 25% |
| 27 | 29,00 | 5,00 | 10,00 | 1,00 | 28,00 | 36,00 | 4,00 | 1,50 | 11,00 | 5,50 | 63% |
| 28 | 15,00 | 1,00 | 3,00 | ,00 | 14,50 | 8,50 | ,50 | ,50 | 1,00 | 6,50 | 67% |
| 29 | 2,00 | 1,00 | 4,50 | ,00 | 1,00 | 3,50 | ,50 | ,00 | 4,50 | 6,00 | 54% |
| 30 | 9,00 | 1,00 | 14,50 | ,00 | 6,00 | 10,00 | ,50 | ,50 | 1,50 | 3,00 | 68% |

[2] In this table the granularity percents are above 30% in most of the cases. At the moment, the authors of these work are studying the sensitivity considering that the granularity is higher or lower than 30%.

Granularity distribution per function type and project, and percent ratio on the total are shown in table 17. Since two different raters measure the same project, the arithmetical mean will be applied.

**Measurement simulation**

In order to determine the static calibration and sensitivity graph in every point for the FP4.1 and EFP4.1 methods, it is necessary to use a simulation process consisting in the measurement of each project with both methods n times, varying in each measurement the count of number of FTRs, RETs, and DETs. This can be done in the following way:

1. Increasing for each function type the number of DETs one by one from 1 to 9 and holding constant the other factors.
2. Increasing in one the number of RETs and FTRs for each data function type and each transactional function type respectively and increasing the number of DETs one by one from 1 to 9 and holding constant the other factors.

## 6. STATIC CALIBRATION AND SENSITIVITY GRAPHS FOR EACH METHOD

First, the static calibration and sensitivity graphs related to the increasing of the count of number of DETs from 1 to 9 for each function type will be obtained. In Figure 1, the static calibration graphs of the FP4.1 method and the EFP4.1 extended method are showed. The count of number of DETs increase is presented in x-axis, and in y-axis the functionality in unadjusted function points increase. The EFP4.1 extended method is printed with ■ and the FP4.1 standard method with ▼.

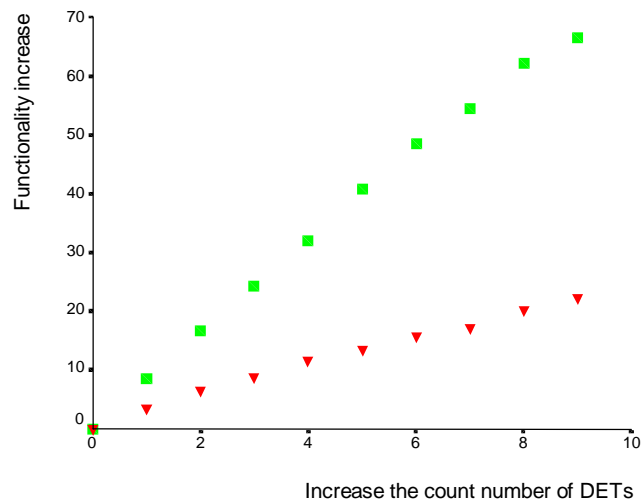

Fig. 1. Static calibration graphs of FP4.1 and EFP4.1 methods

In the EFP4.1 extended method it can be observed that the increase in the number of function points is greater than the FP4.1 standard method when the count of number of DETs grows. These numeric values of the increase are shown in table 18.

In table 18 these numeric incresing values are shown.

Table 18. Numeric values of the functionality increase

| Increment of RETs | Increment of FTRs | Increment of DETs | Functionality increase | |
|---|---|---|---|---|
| | | | FP4.1 | EFP4.1 |
| 0 | 0 | 1 | 3,45 | 8,52 |
| 0 | 0 | 2 | 6,53 | 16,61 |
| 0 | 0 | 3 | 8,70 | 24,27 |
| 0 | 0 | 4 | 11,52 | 32,09 |
| 0 | 0 | 5 | 13,38 | 40,87 |
| 0 | 0 | 6 | 15,73 | 48,48 |
| 0 | 0 | 7 | 17,17 | 54,41 |
| 0 | 0 | 8 | 20,10 | 62,19 |
| 0 | 0 | 9 | 22,22 | 66,55 |

In Figure 2, the sensitivity graphs of the FP4.1 method and the EFP4.1 extended method are showed. The count of number of DETs increase is presented in x-axis, and the sensitivity values in y-axis. The EFP4.1 extended method is printed with ■ and the FP4.1 standard method with ▼.

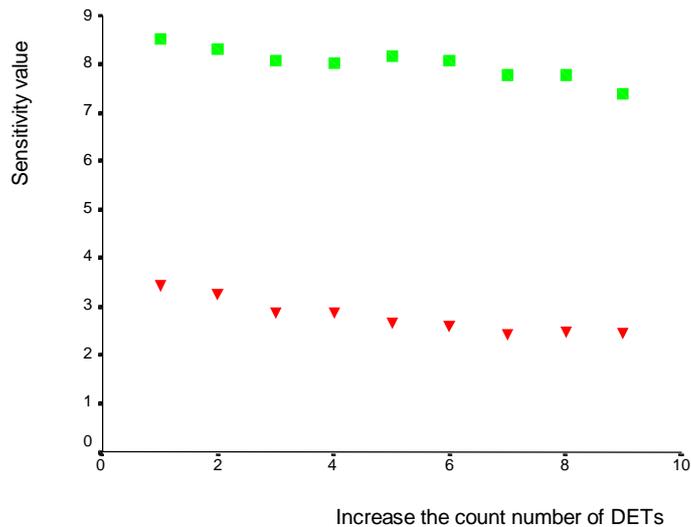

Fig. 2. Sensitivity of FP4.1 and EFP4.1 methods

As it can be observed, the sensitivity in each point is greater in the extended method than in the standard one. Table 19 shows the numeric values of sensitivity in each point.

Table 19. Sensitivity in each point to the FP4.1 and EFP4.1 methods

| Increment of RETs | Increment of FTRs | Increment of DETs | Sensitivity | |
|---|---|---|---|---|
| | | | FP4.1 | EFP4.1 |
| 0 | 0 | 1 | 3,45 | 8,52 |
| 0 | 0 | 2 | 3,27 | 8,30 |
| 0 | 0 | 3 | 2,90 | 8,09 |
| 0 | 0 | 4 | 2,88 | 8,02 |
| 0 | 0 | 5 | 2,68 | 8,17 |
| 0 | 0 | 6 | 2,62 | 8,08 |
| 0 | 0 | 7 | 2,45 | 7,77 |
| 0 | 0 | 8 | 2,51 | 7,77 |
| 0 | 0 | 9 | 2,47 | 7,39 |

Secondly, we will obtain the static calibration and sensitivity graphs corresponding to the increase in one of the count of number of RETs and FTRs for each data function type and each transactional function type respectively, and the increase of the count of number of DETs one by one from 1 to 9 and holding constant the others factors.

In Figure 3, the static calibration graphs of the FP4.1 method and the EFP4.1 extended method are showed. The count of number of DETs[3] increase is presented in x-axis, and the functionality in unadjusted function points increase in y-axis. The EFP4.1 extended method is printed with ■ and the FP4.1 standard method with ▼.

---

[3] Once the RETs and FTRs count is increased in one.

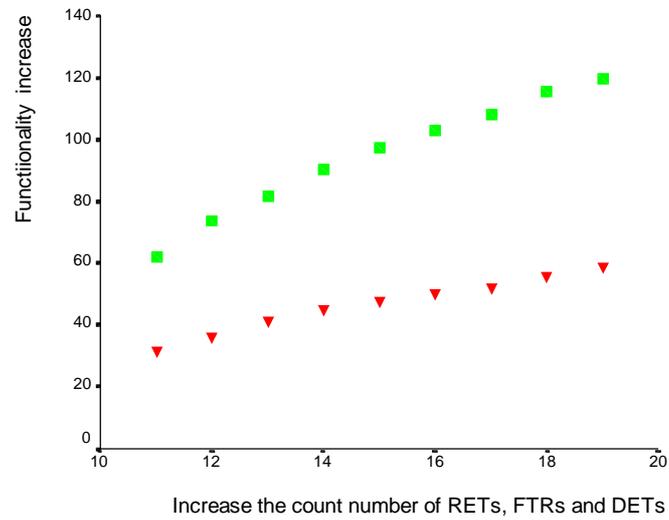

Fig. 3. Static calibration graphs of FP4.1 and EFP4.1 methods

As it can be observed, there is a greater increment in the number of function point in the EFP4.1 extended method, due to the increase in the RETs, FTRs and DETs count number. Table 20 shows the numeric values of these increases.

Table 20: Numeric values of the functionality increase

| Increment of RETs | Increment of FTRs | Increment of DETs | Functionality increase | |
|---|---|---|---|---|
| | | | FP4.1 | EFP4.1 |
| 1 | 1 | 1 | 31,45 | 62,28 |
| 1 | 1 | 2 | 36,27 | 73,49 |
| 1 | 1 | 3 | 41,08 | 81,74 |
| 1 | 1 | 4 | 45,15 | 90,19 |
| 1 | 1 | 5 | 47,53 | 97,52 |
| 1 | 1 | 6 | 50,02 | 103,08 |
| 1 | 1 | 7 | 52,12 | 107,94 |
| 1 | 1 | 8 | 55,77 | 115,58 |
| 1 | 1 | 9 | 59,07 | 119,66 |

In Figure 4, the FP4.1 method and EFP4.1 extended method sensitivity graphs are presented. In x-axis the DETs [3] number increase is presented and in y-axis the sensitivity value.

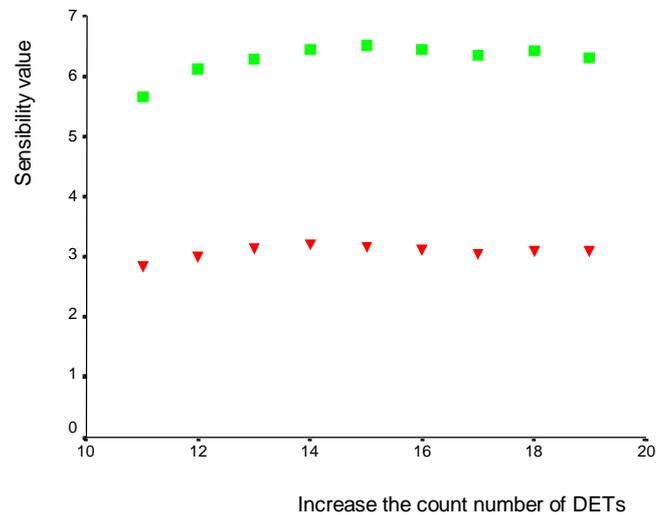

Fig. 4. Sensitivity of FP4.1 and EFP4.1 methods

As it can be observed, the sensitivity in each point is bigger in the EFP4.1 method than in the FP4.1 standard one. Table 21 shows the sensitivity numeric values in each point.

Table 21. Sensitivity in each point to the FP4.1 and EFP4.1 methods

| Increment of RETs | Increment of FTRs | Increment of DETs | Sensitivity | |
|---|---|---|---|---|
| | | | FP4.1 | EFP4.1 |
| 1 | 1 | 1 | 2,86 | 5,66 |
| 1 | 1 | 2 | 3,02 | 6,12 |
| 1 | 1 | 3 | 3,16 | 6,29 |
| 1 | 1 | 4 | 3,23 | 6,44 |
| 1 | 1 | 5 | 3,17 | 6,50 |
| 1 | 1 | 6 | 3,13 | 6,44 |
| 1 | 1 | 7 | 3,07 | 6,35 |
| 1 | 1 | 8 | 3,10 | 6,42 |
| 1 | 1 | 9 | 3,11 | 6,30 |

## CONCLUSIONS

When the difference in the count of number of RETs and DETs or FTRs and DETs is small, the Function Point Analysis method, FPA, does not allow to discriminate the functional size. This is a problem in productivity studies for development projects with a high percentage of granular functions or for the small enhancement projects, since the productivity relates the functional size to the development effort required.

The sensitivity graphs obtained by a simulation process reveal that the Maya and Nguyenkin`s tables incorporated to the EFP4.1 extended method increase its sensitivity in relation to the FP4.1 standard method. Consequently, the EFP4.1 method has a greater power to discriminate the functional size for granular functions.

The static analysis carried out from the data obtained in the measurement of 30 projects with a high granularity level allows us to conclude that there are significant differences between the measurements given by both methods.

It is interesting to divide the projects in two groups depending on the granularity percentage. This is left for a next paper, that will be written in the future.

- Ramón Asensio Monge is with the Department of Computer Science, Oviedo University, C/ Calvo Sotelo s/n  33007 Oviedo, Spain.
  Phone 34 - 98 510 33 70
  Fax    34 - 8 510 33 54
  E-mail: asensio@correo.uniovi.es

- Francisco Sanchis Marco (IEEE Member) is with the Department of Structure and Organization of Information, Politechnical University of Madrid, Km. 7 Carretera Valencia, 28031 Madrid, Spain.
  Phone 34 - 91 336 78 84
  Fax    34 - 1 336 75 20
  E-mail: fsanchis@eui.upm.es

- Fernando Torre Cervigón is with the Department of Computer Science, Oviedo University, C/ Calvo Sotelo s/n  33007 Oviedo, Spain.
  Phone 34 - 98 510 5923
  Fax    34 - 8 510 33 54

  E-mail: torre@lsi.uniovi.es